\documentclass[pre,8pt]{revtex4}
\usepackage[russian, english]{babel}
\usepackage{graphicx}
\usepackage{latexsym}
\usepackage{bbm}
\usepackage{amsmath,amssymb,bm}
\usepackage{cancel}
\addto\captionsrussian{}
\addto\captionsrussian{}

\newcommand{\beq}{\begin{equation}}
\newcommand{\eeq}{\end{equation}}
\newcommand{\ba}{\begin{array}{ccc}}
\newcommand{\ea}{\end{array}}

\def\bea{\begin{eqnarray}}
\def\eea{\end{eqnarray}}
\usepackage[colorlinks=true]{hyperref} 
\hypersetup{
    bookmarks=true,         
    unicode=false,          
    pdftoolbar=true,        
    pdfmenubar=true,        
    pdffitwindow=false,     
    pdfstartview={FitH},    
    pdftitle={My title},    
    pdfauthor={Author},     
    pdfsubject={Subject},   
    pdfcreator={Creator},   
    pdfproducer={Producer}, 
    pdfkeywords={keyword1} {key2} {key3}, 
    pdfnewwindow=true,      
    colorlinks=true,       
    linkcolor=magenta, 
    citecolor=blue,        
    filecolor=magenta,      
    urlcolor=cyan           
}

\renewcommand{\approx}{\simeq}

\renewcommand{\Im}{\text{Im}}

\usepackage{braket}

\begin{document}
\title{The eigenvalue spectrum of a large real antisymmetric random matrix\\ with non-zero mean}
\author{Andrei Katsevich$^{a,b}$, Pavel Meshcheriakov$^{a,b}$}
\affiliation{$^a$Moscow Institute of Physics and Technology, 141700 Dolgoprudny, Russia\\
$^b$Landau Institute for Theoretical Physics, 142432 Chernogolovka, Russia
}

\date{September 3, 2023}
\begin{abstract}
We study the eigenvalue spectrum of a large real antisymmetric random matrix $J_{ij}$. Using a fermionic approach and replica trick, we obtain a semicircular spectrum of eigenvalues when the mean value of each matrix element is zero, and in the case of a non-zero mean, we show that there is a set of critical finite mean values above which eigenvalues arise that are split off from the semicircular continuum of eigenvalues. The result converged with numerical simulations.
\end{abstract}
\maketitle
\section{Introduction}
For even $N$ consider a purely imaginary antisymmetric $N\times N$ matrix $iJ_{ij}=-iJ_{ji}.$ Its eigenvalues are decomposed naturally into pairs $\pm \lambda_{i}\in\mathbb{R}$ with $i=1,2,\dots, N/2$ and $\lambda_{i}> 0$. We define the density of eigenvalues as 
\begin{equation}
    \nu(\lambda) = \frac{1}{N}\sum_{i=1}^{N/2}\delta(\lambda-\lambda_{i})+\delta(\lambda+\lambda_{i})\,.
\end{equation}
We would like to find $\nu(\lambda)$ in the large $N$ limit for the random antisymmetric matrix $J_{ij}$ whose elements are taken from Gaussian ensemble with fixed mean $J_{0}/N$ and variance $\sigma^{2}=J^{2}/N$ ($J_0$ and $J$ are of the order of unity), so the probability density function reads
\begin{equation}
    p(J_{ij}) = \frac{1}{\sqrt{2\pi \sigma^{2}}} \exp\left(-\frac{(J_{ij}-J_{0}/N)^{2}}{2\sigma^{2}}\right)\,.
\end{equation}

The eigenvalue probability density function for some ensembles of Gaussian random matrices with zero mean $J_0=0$ is a Wigner semicircle. This result was obtained in classical works on various matrix ensembles. See, for example, the book \cite{Meh2004}. Several examples include Gaussian orthogonal (GOE), unitary (GUE), and symplectic (GSE) ensembles.\\
For these ensembles, it was showed that the eigenvalue probability density function of the matrix ensemble approaches a Wigner semicircle as the size of the matrices increases to infinity. This is a fundamental result in random matrix theory, demonstrating the universality of the Wigner semicircle.
\begin{equation}
    \rho_0(\lambda)\equiv\braket{\nu(\lambda)}_{J_{ij}:J_0=0}=
    \begin{cases}
        \begin{aligned}
            &\frac{1}{\pi J}\sqrt{1-\frac{\lambda^2}{4J^2}}\,, &\quad &|\lambda|<2J\,;\\
            &0\,, &\quad &|\lambda|>2J\,.
        \end{aligned}
    \end{cases}
\end{equation}
In the case of antisymmetric matrices Wigner semicircle was obtained in \cite{MEHTA1968449}. Our goal is to generalize this result to arbitrary $J_0\neq0$.

A similar problem in the case of real symmetric random matrices $\bm{M}$ with mean $\frac{M_0}{N}$ was discussed in \cite{Edwards_1976, Jones_1978}. Using replica trick, it was shown that if the mean value of random matrix elements is greater than a critical one the distribution of the eigenvalues consists of the Wigner semicircle plus a separate single state: 
\begin{equation}
    \rho_{M_0}(\lambda)\equiv\braket{\nu(\lambda)}_{M_{ij}:M_0\neq0}=
    \begin{cases}
        \begin{aligned}
            &\rho_0(\lambda)+\frac{1}{N}\delta\left(\lambda-M_0-\frac{J^2}{M_0}\right)\,,&\quad &|M_0|>J\,;\\
            &\rho_0(\lambda)\,, &\quad &|M_0|<J\,.
        \end{aligned}
    \end{cases}
\end{equation}

An algorithm for applying the replica trick:
\begin{enumerate}
    \item Rewrite delta-functions in $\nu(\lambda)$ as derivatives of logarithmic functions using Sokhotski–Plemelj formula.
    \item Rewrite logarithmic function in $\nu(\lambda)$ using limit:
    \begin{equation}
        \log x=\lim\limits_{n\rightarrow0}\frac{x^n-1}{n}\,.
    \end{equation}
    \item Swap the limit $n\rightarrow0$ with the limit $N\rightarrow\infty$. Treat $n$ as an integer during calculating the Gaussian integrals.
    \item Take a limit $n\rightarrow0$ after evaluating the integrals.
\end{enumerate}

Physically, the replica trick introduces $n$ replicated copies of the original disordered system that are coupled together. It uses the non-commutativity of limits $n\rightarrow0$ and $N\rightarrow\infty$. Taking the limit $n \rightarrow 0$ at the end serves to analytically continue the results for integer $n$ to non-integer values. The coupling between the replicated systems enables averaging over the disorder in the original system.

This problem also arises when one considers the $\text{SYK}_2$ model with $\frac{N(N-1)}{2}$ couplings \cite{KlebTarn}:
\begin{equation}
    H_{SYK}=i\sum_{i<j}J_{ij}\psi_i\psi_j,
\end{equation}
where $\psi_i, i = 1,...,N$ are Majorana fermions with anticommutation rules $\{\psi_i,\psi_j\}=\delta_{ij}$ and the random couplings $J_{ij}=-J_{ji}$ are drawn from Gaussian distribution with non-zero mean $J_0/N$ and variance $\sigma^2=J^2/N$:
\begin{equation}
   \braket{J_{ij}}=\frac{J_0}{N}, \quad \braket{(J_{ij}-J_0/N)^2}=\frac{J^2}{N}\,. 
\end{equation}
As in the case of a random matrix, $\text{SYK}_2$ is likely to have similar properties.

\section{The technique}
In this section we give a general scheme of calculations. The presented formulas can be applied to an arbitrary matrix $\bm{H}$, so here we do not restrict ourselves to the case of antisymmetric ones. We can write $\nu(\lambda)$ as 
\begin{equation}
    \nu(\lambda) = \frac{1}{\pi N}\sum_{i=1}^{N}\textrm{Im}\frac{1}{\lambda-i\epsilon - \lambda_{i}}\,,
\end{equation}
where we used Sokhotski–Plemelj formula
\begin{equation}
\lim_{\epsilon\to 0^{+}}\frac{1}{x-i\epsilon} = i\pi \delta(x) + \mathcal{P}\Big(\frac{1}{x}\Big)\,.
\end{equation}
On the other hand we have
\begin{equation}
    \textrm{det}(\bm{I}\lambda-\bm{H})= \prod_{i=1}^{N}(\lambda-\lambda_i)\,.
\end{equation}
Therefore, we obtain
\begin{equation}\label{nu}
    \nu(\lambda) = \frac{1}{\pi N}\textrm{Im} \frac{\partial}{\partial\lambda} \ln \textrm{det}(\bm{I}(\lambda-i\epsilon)-\bm{H})
\end{equation}
and using replicas
\begin{equation}
    \nu(\lambda) = \frac{1}{\pi N}\text{Im} \frac{\partial}{\partial\lambda} \lim_{n\to 0} \frac{1}{n}\{\det(\bm{I}\lambda_{\epsilon}-\bm{H})^n-1\}\,,
\end{equation}
where $\lambda_{\epsilon} \equiv \lambda - i\epsilon$. The determinant of a matrix may be represented as an integral over the Grassmann numbers (fermionic approach):
\begin{equation}
    \textrm{det}(\bm{I}\lambda_{\epsilon}-\bm{H}) = \int \prod_{i=1}^Nd\bar{c}_{i} dc_{i}\exp\left(-\sum\limits_{i,j}\bar{c}_i(I_{ij}\lambda_{\epsilon}-H_{ij})c_j\right)\,,
\end{equation}
where $\{\bar{c}_{i},c_{j}\} =\{\bar{c}_{i},\bar{c}_{j}\} = \{c_{i},c_{j}\}=0$. Thus, we obtain the basic working formula:
\begin{equation}\label{spectralDensity}
    \nu(\lambda)=\frac{1}{\pi N}\textrm{Im} \frac{\partial}{\partial\lambda} \lim_{n\to 0} \frac{1}{n} \left\{\int \prod_{i;\alpha}d \bar{c}^\alpha_{i} dc^\alpha_{i}\exp\left(-\sum\limits_{i,j;\alpha}\bar{c}^\alpha_{i}(I_{ij}\lambda_{\epsilon}-H_{ij})c^\alpha_{j}\right)-1 \right\}\,.
\end{equation}
\section{Spectrum of the matrix with zero mean}
In this section we elaborate on the case of antisymmetric matrices with zero mean. This is important because the case of non-zero mean will refer to the case of zero mean. Now we consider an antisymmetric matrix in which each element has zero mean with the probability density function:
\begin{equation}
    p(J_{ij})=\frac{1}{\sqrt{2\pi\sigma^2}}\exp\left(-\frac{J_{ij}^2}{2\sigma^{2}}\right)\,.
\end{equation}
The average density of eigenvalues:
\begin{equation}\label{rho0:0}               
    \rho_0(\lambda)=\braket{\nu(\lambda)}_{J_{ij}:J_0=0}= \int \nu\left(\lambda;\{J_{ij}:J_0=0\}\right)\prod\limits_{i<j}\; p(J_{ij})\;dJ_{ij}\,.
\end{equation}
Using (\ref{spectralDensity}) and (\ref{rho0:0}), we obtain
\begin{multline}\label{rho0:1}
    \rho_0(\lambda)=\frac{1}{\pi N}\textrm{Im}\frac{\partial}{\partial \lambda}\lim\limits_{n\rightarrow0}\frac{1}{n}\Biggl\{\int\left(\prod\limits_{i;\alpha}d \bar{c}_i^{\alpha}d c_i^{\alpha}\right)\exp\left(-\lambda_{\epsilon}\sum\limits_{i;\alpha}\bar{c}_i^\alpha c_i^\alpha\right)\times
    \\\times\left(\prod\limits_{i<j}\frac{dJ_{ij}}{2\pi\sigma^{2}}\exp\left(-\frac{J_{ij}^2}{2\sigma^{2}}\right)\right)\exp\left(i\sum\limits_{i,j;\alpha}\bar{c}_i^\alpha J_{ij}c^\alpha_j\right)-1\Biggl\}\,.
\end{multline}

Let it be $P_{ij}= \sum\limits_\alpha\bar{c}^\alpha_i c^\alpha_j$ is antihermitian tensor ($P_{ij}=-\bar{P}_{ji}$) and $P^A_{ij}=\sum\limits_\alpha\frac{\bar{c}^\alpha_ic^\alpha_j-\bar{c}^\alpha_jc^\alpha_i}{2}$ is its antisymmetric part. Extract and carry out the Gaussian integrals over $J_{ij}$:
\begin{equation}\label{GaussianI1}
    I_1 = \frac{1}{(2\pi\sigma^2)^{\frac{N(N-1)}{2}}}\int \left(\prod\limits_{i<j}dJ_{ij}\right)\exp\left(\sum\limits_{i,j}iP_{ij}J_{ij}-\sum\limits_{i<j}\frac{J_{ij}^2}{2\sigma^{2}}\right)=\exp\left(-\sigma^2\sum\limits_{i,j}(P^A_{ij})^2\right)\,.
\end{equation}
Substituting (\ref{GaussianI1}) into (\ref{rho0:1}), we obtain
\begin{equation}\label{rho0:2}
    \rho_0(\lambda)=\frac{1}{\pi N}\textrm{Im}\frac{\partial}{\partial \lambda}\lim\limits_{n\rightarrow0}\frac{1}{n}\Biggl\{\int\left(\prod\limits_{i;\alpha}d \bar{c}_i^{\alpha}d c_i^{\alpha}\right)\exp\left(-\lambda_{\epsilon}\sum\limits_{i}P_{ii}-\sigma^2\sum\limits_{i, j}(P^A_{ij})^2\right)-1\Biggl\}\,.
\end{equation}

Let us define the antihermitian tensor $T^{\alpha\beta}=\sum\limits_i\bar{c}_{i}^{\alpha}c_{i}^{\beta}$ ($T^{\alpha\beta}=-\bar{T}^{\beta\alpha}$) with symmetric and antisymmetric parts $T_S^{\alpha\beta}=\sum\limits_i\frac{\bar{c}_i^\alpha c_i^\beta+\bar{c}_i^\beta c_i^\alpha}{2}$, $T_A^{\alpha\beta}=\sum\limits_i\frac{\bar{c}_i^\alpha c_i^\beta-\bar{c}_i^\beta c_i^\alpha}{2}$ and antisymmetric tensor $K^{\alpha\beta}=\sum\limits_{i} c_{i}^{\alpha}c_{i}^{\beta}$ ($K^{\alpha\beta}=-K^{\beta\alpha}$). We rewrite the tensor expressions associated with $\bm{P}, \bm{P}^A$ in terms of $\bm{T}, \bm{K}$ in (\ref{rho0:2}). Thus, we obtain
\begin{equation}\label{rho0:3}
    \rho_0(\lambda)=\frac{1}{\pi N}\textrm{Im}\frac{\partial}{\partial \lambda}\lim\limits_{n\rightarrow0}\frac{1}{n}\Biggl\{\int\left(\prod\limits_{i;\alpha}d \bar{c}_i^{\alpha}dc_i^{\alpha}\right)\exp\left(-\lambda_{\epsilon}\sum\limits_\alpha T^{\alpha\alpha}-\frac{J^2}{2N}\sum\limits_{\alpha,\beta}\left(T^{\alpha\beta}T^{\beta\alpha}+\bar{K}^{\alpha\beta}K^{\beta\alpha}\right)\right)-1\Biggl\}\,.
\end{equation}

For convenience, we introduce 
\begin{equation}\label{I2:1}
    I_2=\int\left(\prod\limits_{i;\alpha}
    d\bar{c}_i^\alpha dc_i^\alpha\right) \exp\left(-\lambda_{\epsilon}\sum\limits_{\alpha}T^{\alpha\alpha}-\frac{J^2}{2N}\sum_{\alpha,\beta}\left( T^{\alpha\beta}T^{\beta \alpha} +\bar{K}^{\alpha\beta}K^{\beta\alpha}\right)\right)\,.
\end{equation}
The next step is to apply the Hubbard-Stratonovich transformation to the tensors $\bm{T}, \bm{K}$, which allows us to replace quadratic expressions with tensors with linear ones by introducing additional integration variables. For $\bm{T}$ we have
\begin{multline}\label{TTHubburd}
    \exp\left(-\frac{J^2}{2N}\sum\limits_{\alpha,\beta}T^{\alpha\beta}T^{\beta\alpha}\right)=\exp\left(-\frac{J^2}{2N}\text{tr}\;\bm{T}^2\right)=\exp\left(-\frac{J^2}{2N}\text{tr}\;\bm{T}_S^2-\frac{J^2}{2N}\text{tr}\;\bm{T}_A^2\right)=
    \\=2^{-\frac{n}{2}}\left(\frac{N\lambda^2}{\pi J^2}\right)^{\frac{n^2}{2}}\int[d\bm{R}_S][d\bm{R}_A]\exp\left(-\frac{N\lambda^2}{2J^2}\text{\;tr\;}\bm{R}_S^2+\frac{N\lambda^2}{2J^2}\text{\;tr\;}\bm{R}_A^2-i\lambda\text{\;tr\;}\bm{T}_S\bm{R}_S+\lambda \text{\;tr\;}\bm{T}_A\bm{R}_A\right)\,,
\end{multline}
where $\bm{R}_S$ is a real symmetric $\bm{R}_S^T=\bm{R}_S$ and $\bm{R}_A$ is a real antisymmetric $\bm{R}_A^T=-\bm{R}_A$ matrices.
\begin{equation}
    [d\bm{R}_S]=\prod\limits_{\alpha\leq\beta}dR_S^{\alpha\beta},\quad[d\bm{R}_A]=\prod\limits_{\alpha<\beta}dR_A^{\alpha\beta}\,.
\end{equation}

For the tensor $\bm{K}$ we have:
\begin{multline}\label{KbarKHubburd}
    \exp\left(-\frac{J^2}{2N}\sum\limits_{\alpha,\beta}\bar{K}^{\alpha\beta}K^{\beta\alpha}\right)=\exp\left(-\frac{J^2}{2N}\text{tr\;} \bar{\bm{K}}\bm{K}\right)=
    \\=2^{n(n-1)}\left(\frac{N\lambda^2}{\pi J^2}\right)^{\frac{n(n-1)}{2}}\int[d\bm{M}]\exp\left(\frac{2N\lambda^2}{J^2}\text{\;tr\;} \bar{\bm{M}}\bm{M}+\lambda\text{\;tr\;} \bm{K}\bm{M}+\lambda\text{\;tr\;} \bar{\bm{K}}\bar{\bm{M}}\right)\,,
\end{multline}
where $\bm{M}$ is an antisymmetric complex matrix and $[d\bm{M}]=\prod\limits_{i<j}d^2M_{ij}$. Substitute (\ref{TTHubburd}) and (\ref{KbarKHubburd}) in (\ref{I2:1}) and obtain
\begin{multline}\label{I2:2}
    I_2=2^{n(n-\frac{3}{2})}\left(\frac{N\lambda^2}{\pi J^2}\right)^{n(n-\frac{1}{2})}\int\left(\prod\limits_{i;\alpha} d\bar{c}_{i}^{\alpha}dc_{i}^\alpha\right) [d\bm{R}_S][d\bm{R}_A][d\bm{M}]\exp\left(-\lambda_\epsilon\text{tr}\;\bm{T}-\frac{N\lambda^2}{2J^2}\text{tr\;}\bm{R}_S^2+\frac{N\lambda^2}{2J^2}\text{tr\;}\bm{R}_A^2-\right.\\\left.-i\lambda\text{\;tr\;}\bm{T}_S\bm{R}_S+\lambda \text{\;tr\;}\bm{T}_A\bm{R}_A+\frac{2N\lambda^2}{J^2}\text{tr}\;\bar{\bm{M}}\bm{M}+\lambda\text{\;tr\;}\bm{K}\bm{M}+\lambda\text{\;tr\;} \bar{\bm{K}}\bar{\bm{M}}\right)\,.
\end{multline}

Let us introduce the vector of $2n$ parameters $\bm{\theta}_i=\{c^1_i,...,c^n_i,\bar{c}^1_i,...,\bar{c}^n_i\}$ with the measure $ d\bm{\theta}_i=\prod\limits_\alpha d\bar{c}_i^\alpha dc_i^\alpha$ and we represent the linear tensor part in a more compact form
\begin{equation}\label{linearTensorPart1}
    -\lambda_\epsilon\text{tr}\;\bm{T}-i\lambda\text{\;tr\;}\bm{T}_S\bm{R}_S+\lambda\text{\;tr\;}\bm{T}_A\bm{R}_A+\lambda\;\text{tr}\;\bm{K}\bm{M}+\lambda\text{\;tr\;} \bar{\bm{K}}\bar{\bm{M}}=-\sum_i\frac{1}{2}\bm{\theta}^{T}_i\left(\lambda_\epsilon \bm{\Omega}+i\lambda\bm{\sigma}_1\bm{L}\right)\bm{\theta}_i\,,
\end{equation}
where $\bm{R}=\frac{\bm{R}_S+i\bm{R}_A}{2}$ is the Hermitian matrix, $\bm{\sigma}_1$ is the Pauli matrix and also
\begin{equation}
    \bm{\Omega}=
    \begin{pmatrix}
        \bm{0} & -\bm{I}_n\\\
       \bm{I}_n & \bm{0}
    \end{pmatrix},
    \quad \bm{L}=2
    \begin{pmatrix}
       \bm{R}^T & -i\bar{\bm{M}}\\
        -i\bm{M} & -\bm{R}
    \end{pmatrix}\,.
\end{equation}
It should be noted that the matrix $\bm{L}$ has the following properties
\begin{equation}
     \bm{L}=\bm{L}^{\dagger},\quad \bm{\sigma}_1\bm{L}\bm{\sigma}_1=-\bm{L}^T\,.
\end{equation}
In terms of the Cartan classification of Hamiltonians, we work with class $D$ with $P^2=1$ symmetry (see appendix \ref{Cartan}). This class was considered, for example, in \cite{2002} using a supersymmetric approach.

The next step is to perform integration over the Grassmannian variables in (\ref{I2:2}). To do this, we use the formula for the Berezin integral:
\begin{equation}
    \int d\bm{\theta}_i\exp{\left(-\frac{1}{2}\bm{\theta}^{T}_i\left(\lambda_\epsilon\bm{\Omega}+i\lambda\bm{\sigma}_1\bm{L}\right)\bm{\theta}_i\right)}=\text{Pf}(\lambda_\epsilon\bm{\Omega}+i\lambda\bm{\sigma}_1\bm{L})\,.
\end{equation}
We obtain
\begin{equation}\label{I2:3}
    I_2=c_n\int[d\bm{R}][d\bm{M}]\exp\left(-\frac{2N\lambda^2}{J^2}\text{tr}\left(\bm{R}^2-\bar{\bm{M}}\bm{M}\right)+\frac{N}{2}\log\det\left(\lambda_{\epsilon}\bm{\Omega}+i\lambda\bm{\sigma}_1\bm{L}\right)\right),
\end{equation}
where $c_n=2^{n(n-\frac{3}{2})}\left(\frac{N\lambda^2}{\pi J^2}\right)^{n(n-\frac{1}{2})}$. We keep in mind that
\begin{equation}
    \text{tr}\;\bm{L}^2=8\text{tr}\;(\bm{R}^2-\bar{\bm{M}}\bm{M}),\quad \bm{\Omega}^2=-\bm{I},\quad \bm{\Omega}\bm{\sigma}_1=
    \begin{pmatrix}
        -\bm{I}_n & 0\\
        0 & \bm{I}_n
    \end{pmatrix}=-\bm{\sigma}_3\,.
\end{equation}
Therefore, we can rewrite (\ref{I2:3}) through one matrix $\bm{L}$ over which the integration is performed
\begin{equation}\label{I2:4}
    I_2=c_n\int[d\bm{L}]\exp\left(-N\text{tr}\left(\frac{\lambda^2}{4J^2}\bm{L}^2-\frac{1}{2}\log\left(\lambda_\epsilon\bm{I}+i\lambda\bm{\sigma}_3\bm{L}\right)\right)\right)\,.
\end{equation}
By varying the action, the equation of the saddle-point manifold is 
\begin{equation}
    \frac{\lambda^2}{2J^2}\bm{L}_0-\frac{1}{2}i\lambda\bm{\sigma}_3(\lambda_\epsilon\bm{I}+i\lambda\bm{\sigma}_3\bm{L}_0)^{-1}=0\,,
\end{equation}
and it is solved by
\begin{equation}
    \bm{L}_0=\text{diag}(x_1,...,x_{n},x_{n+1},...,x_{2n})\,,
\end{equation}
where $x_i$ are saddle points
\begin{equation}
    \begin{cases}
        x^{\pm}_i=\frac{i\lambda_\epsilon\pm\sqrt{4J^2-\lambda_\epsilon^2}}{2\lambda},\qquad i=\{1,...,n\}\,;\\
        x^{\pm}_j=-\frac{i\lambda_\epsilon\pm\sqrt{4J^2-\lambda_\epsilon^2}}{2\lambda},\quad j=\{n+1,...,2n\}\,.
    \end{cases}
\end{equation}

Let us introduce $g_{\pm}(x)$ as
\begin{equation}
    g_{\pm}(x)=\frac{\lambda^2}{4J^2}x^2-\frac{1}{2}\log\left(\lambda_\epsilon\pm ix\lambda\right)\,.
\end{equation}
By deforming the integration contour onto the saddle-point manifold, the integral (\ref{I2:4}) reduces to
\begin{equation}
    I_2\approx c_n\prod\limits_{i=1}^{n}\left(\sum_{x_i\in\{x_i^+,x_i^-\}}\int_{-\infty}^\infty dx_i e^{-Ng_+(x_i)}\right)\prod\limits_{j=n+1}^{2n}\left(\sum_{x_j\in\{x_j^+,x_j^-\}}\int_{-\infty}^\infty dx_je^{-Ng_-(x_j)}\right)\,.
\end{equation}
Considering $g_{+}(x)=g_{-}(-x)\equiv g(x)$, we get
\begin{equation}
    I_2\approx c_n\prod\limits_{i=1}^{n}\left(\sum_{x_i\in\{x_i^+,x_i^-\}}\int_{-\infty}^\infty dx_i e^{-Ng_+(x_i)}\right)^2\,.
\end{equation}

Without loss of generality, let it be $\lambda>0$. We know that $\rho_0(\lambda)=\rho_0(-\lambda)$ due to the matrix antisymmetry.
\begin{figure}
    \centering
    \includegraphics[scale=0.7]{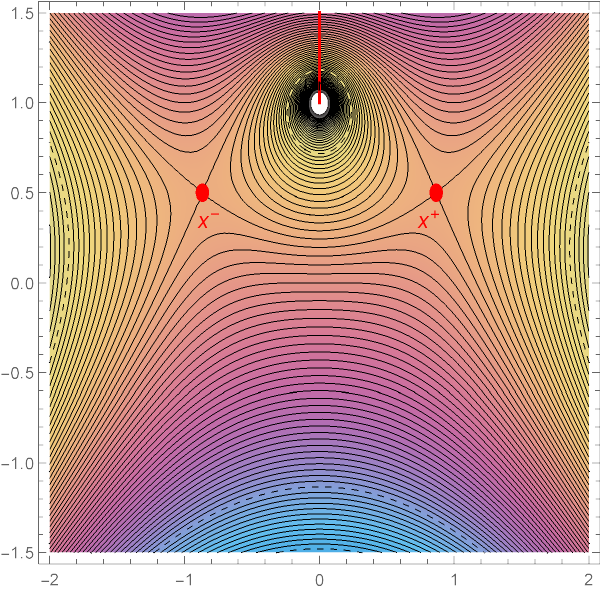}
    \caption{Here $x^{\pm}$ are the saddle points of $g(x)$. We choose only $x^-$. Red lines means branch cut.} 
    \label{Contour}
\end{figure}
We investigate 2 cases: 
\begin{itemize}
    \item $\lambda<2J$.

    Value of $g(x)$ in $x_i^\pm$:
    \begin{equation}
        g(x^\pm_i)=\frac{1}{4}+\frac{\lambda_\epsilon^2}{8J^2}\left(-1\pm i\sqrt{\frac{4J^2}{\lambda_\epsilon^2}-1}\right)-\frac{1}{2}\log\left(\frac{\lambda_\epsilon}{2}\left(1\pm i\sqrt{\frac{4J^2}{\lambda_\epsilon^2}-1}\right)\right)\,.
    \end{equation}
    In the limit $\epsilon\rightarrow0$:
    \begin{equation}
        g''(x^\pm_i)=\frac{\lambda^4}{4J^4}\left(\frac{4J^2}{\lambda^2}-1\pm i\sqrt{\frac{4J^2}{\lambda^2}-1}\right)\,,
    \end{equation}
    \begin{equation}
        |g''(x^\pm_i)|=\frac{\lambda^2}{2J^3}\sqrt{4J^2-\lambda^2},\quad \arg g''(x^\pm_i)=\pm\arctan\left(\frac{1}{\sqrt{\frac{4J^2}{\lambda^2}-1}}\right)\,.
    \end{equation}
    For all $i$ there are two variants: $x_i=x^-_i$ or $x_i=x^+_i$. Therefore, we have a $n$-dimensional complex saddle-point manifold. We choose all $x_i=x^-_i\equiv x^-$ to $\rho_0(\lambda)>0$ because contributions of all other variants are suppressed (see fig. \ref{Contour}). Similar reasoning was made in \cite{Kamenev_1999} for GUE, which consists of Hermitian matrices.
    
    Since $N\gg1$, we use the complex steepest descent method:
    \begin{equation}
        \rho_0(\lambda)\approx\frac{1}{\pi N}\text{Im}\frac{\partial}{\partial \lambda}\log\left(\sqrt{\frac{2\pi}{N|g''(x^-)|}}e^{-Ng(x^-)-i\frac{\arg g''(x^-)}{2}}\right)^2\,.
    \end{equation}
    As the matrix size $N$ grows to infinity, $|g''(x^-_i)|$ and $\arg g''(x^-_i)$ remain approximately constant, scaling as $\mathcal{O}(1)$ for each $i$. We derive
    \begin{equation}
        \rho_0(\lambda)\approx\frac{1}{\pi}\text{Im}\frac{\partial}{\partial\lambda}(-2g(x^-))=\frac{1}{\pi J}\sqrt{1-\frac{\lambda^2}{4J^2}}\,.
    \end{equation}
    \item $\lambda>2J$.\\
    Both saddle points are purely imaginaries:
    \begin{equation}
        x^\pm_i=\frac{i}{2}\left(1\pm \sqrt{1-\frac{4J^2}{\lambda^2}}\right)\,.
    \end{equation}
    We have $g(x^\pm_i)\in\mathbb{R}$, consequently $\rho_0(\lambda)=0$.
\end{itemize}
    
\begin{figure}
    \centering
    \includegraphics[scale=0.7]{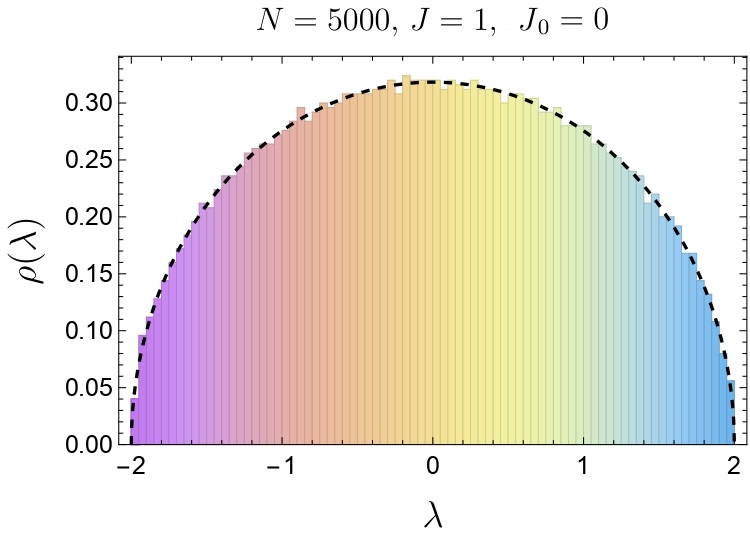}
    \caption{Wigner semicircle. The dotted line corresponds $\rho_0(\lambda).$ }
    \label{Wigner1}
\end{figure}

Combining the results of both cases and remembering that $\lambda$ can be negative, we get
\begin{equation}
    \rho_0(\lambda)=
    \begin{cases}
        \begin{aligned}
            &\frac{1}{\pi J}\sqrt{1-\frac{\lambda^2}{4J^2}}, &\quad &|\lambda|<2J\,;\\
            &0, &\quad &|\lambda|>2J\,.
        \end{aligned}
    \end{cases}
\end{equation}
This is the Wigner semicircle (see fig. \ref{Wigner1}). We reproduce this well known result for the $D$ class using a replica approach with fermions.
    
\section{Spectrum of the antisymmetric matrix with non-zero mean}
In this section we consider antisymmetric matrices with elements having non-zero mean $\frac{J_0}{N}$ with the probability density function:
\begin{equation}
    p(J_{ij}) =\frac{1}{\sqrt{2\pi \sigma^{2}}} e^{-\frac{(J_{ij}-J_0/N)^{2}}{2\sigma^{2}}}\,.
\end{equation}
The average density of eigenvalues:
\begin{equation}
    \rho_{J_0}(\lambda)=\braket{\nu(\lambda)}_{J_{ij}:J_0\neq0}= \int \nu\left(\lambda;\{J_{ij}:J_0\neq0\}\right)\prod\limits_{i<j}\; p(J_{ij})\;dJ_{ij}\,.
\end{equation}
Similarly to (\ref{rho0:1}) we have
\begin{multline}
    \rho_{J_0}(\lambda)=\frac{1}{\pi N}\textrm{Im}\frac{\partial}{\partial \lambda}\lim\limits_{n\rightarrow0}\frac{1}{n}\Biggl\{\int\left(\prod\limits_{i;\alpha}d \bar{c}_{i}^{\alpha}d c_{i}^{\alpha}\right)\exp\left(-\sum\limits_{i,j;\alpha}\bar{c}_i^\alpha(\lambda_{\epsilon}I_{ij}-iJ_{ij})c^\alpha_j\right)\times
    \\\times\left(\prod\limits_{i<j}\frac{dJ_{ij}}{\sqrt{2\pi\sigma^2}}\exp\left(-\frac{(J_{ij}-\frac{J_0}{N})^2}{2\sigma^2}\right)\right)-1\Biggl\}\,.
\end{multline}

Now we are reducing the problem with a non-zero mean to a zero-mean case. For this purpose, add and subtract term $A_{ij}\frac{J_0}{N}$ to $J_{ij}$:
\begin{equation}
    J_{ij}=\underbrace{J_{ij}-A_{ij}\frac{J_0}{N}}_{\text{zero-mean matrix}}+A_{ij}\frac{J_0}{N}\,,
\end{equation}
where $A_{ij}$ is a constant antisymmetric matrix (see appendix \ref{ConstantMatrix}). Using the result of (\ref{rho0:3}), we get
\begin{multline}\label{rhoJ0}
    \rho_{J_0}(\lambda) = \frac{1}{\pi N}\textrm{Im} \frac{\partial}{\partial \lambda} \lim_{n\to 0} \frac{1}{n}\Biggl\{ \int\left(\prod\limits_{i;\alpha}d \bar{c}_i^{\alpha}d c_i^{\alpha}\right) \exp\bigg(-\sum\limits_{i,j;\alpha}\bar{c}^\alpha_{i}\left(\bm{I}\lambda_{\epsilon}-i\bm{A}\frac{J_0}{N}\right)_{ij}c^{\alpha}_{j}-
    \\-\frac{J^2}{2N}\sum_{\alpha,\beta}\left(T^{\alpha\beta}T^{\beta\alpha}+\bar{K}^{\alpha\beta}K^{\beta\alpha}\right)\bigg)-1 \Biggl\}\,.
\end{multline}
As in the case of the zero mean for the integral in (\ref{rhoJ0}), we perform the Hubbard-Stratonovich transformation and obtain an analogous formula to (\ref{I2:2}):
\begin{multline}\label{I2J0}
    I_2=c_n\int\left(\prod\limits_{i;\alpha} d\bar{c}_{i}^{\alpha}dc_{i}^{\alpha}\right) [d\bm{R}_S][d\bm{R}_A][d\bm{M}]\exp\Biggl(-\sum\limits_{i,j;\alpha}\bar{c}^\alpha_{i}\left(\lambda_{\epsilon}\bm{I}-i\bm{A}\frac{J_0}{N}\right)_{ij}c^\alpha_{j}-\frac{N\lambda^2}{2J^2}\text{tr\;}\bm{R}_S^2+\frac{N\lambda^2}{2J^2}\text{tr\;}\bm{R}_A^2-
    \\-i\lambda\text{\;tr\;}\bm{T}_S\bm{R}_S+\lambda \text{\;tr\;}\bm{T}_A\bm{R}_A+\frac{2N\lambda^2}{J^2}\text{tr}\;\bar{\bm{M}}\bm{M}+\lambda\text{\;tr\;}\bm{K}\bm{M}+\lambda\text{\;tr\;} \bar{\bm{K}}\bar{\bm{M}}\Biggl)\,.
\end{multline}
Performing calculations for the linear part in (\ref{I2J0}) similarly to (\ref{linearTensorPart1}), we come to 
\begin{multline}
    -\sum\limits_{i,j;\alpha}\bar{c}^\alpha_{i}\left(\lambda_{\epsilon}\bm{I}-i\bm{A}\frac{J_0}{N}\right)_{ij}c^\alpha_{j}-i\lambda\text{\;tr\;}\bm{T}_S\bm{R}_S+\lambda \text{\;tr\;}\bm{T}_A\bm{R}_A+\frac{2N\lambda^2}{J^2}\text{tr}\;\bar{\bm{M}}\bm{M}+\lambda\text{\;tr\;}\bm{K}\bm{M}+\lambda\text{\;tr\;} \bar{\bm{K}}\bar{\bm{M}}=
    \\=\sum\limits_{i,j}\bm{A}^{ij}\otimes\frac{1}{2}\bm{\theta}^T_i\left(\frac{iJ_0}{N}\bm{\sigma}_1\right)\bm{\theta}_j  -\sum_{i,j}\bm{I}^{ij}\otimes\frac{1}{2}\bm{\theta}^{T}_i\left(\lambda_\epsilon \bm{\Omega}+i\lambda\bm{\sigma}_1\bm{L}\right)\bm{\theta}_j\,,
\end{multline}
where $\bm{A}^{ij}=\bm{A}$ and $\bm{I}^{ij}=\bm{I}$ for all $i$ and $j$. After integration by all $\bm{\theta}_i$, we get a formula similar to (\ref{I2:3}):
\begin{equation}
    I_3=c_n\int[d\bm{R}][d\bm{M}]\exp\left(-\frac{2N\lambda^2}{J^2}\text{tr}\left(\bm{R}^2-\bar{\bm{M}}\bm{M}\right)+\frac{N}{2}\log\text{det}\left(\bm{A}\otimes\frac{iJ_0}{N}\bm{\sigma}_1 -\bm{I}\otimes\left(\lambda_\epsilon\bm{\Omega}+i\lambda\bm{\sigma}_1\bm{L}\right)\right)\right)\,.
\end{equation}
Later we diagonalize $\bm{A}$:
\begin{equation}
    \text{det}\left(\bm{A}\otimes\frac{iJ_0}{N}\bm{\sigma}_1 -\bm{I}\otimes\left(\lambda_\epsilon \bm{\Omega}+i\lambda\bm{\sigma}_1\bm{L}\right) \right)=\text{det}\left(\bm{D}\otimes\frac{iJ_0}{N}\bm{\sigma}_1-\bm{I}\otimes\left(\lambda_\epsilon \bm{\Omega}+i\lambda\bm{\sigma}_1\bm{L}\right) \right)\,,
\end{equation}
where $\bm{D}=\text{diag}(i\lambda_1,...,i\lambda_{N})$ (see appendix \ref{ConstantMatrix}). After substitution of $\bm{D}$, we obtain
\begin{equation}
    I_3=c_n\int[d\bm{L}]\prod_{i=1}^{N}\text{det}^{\frac{1}{2}}\left(\bm{I}+\frac{J_0\lambda_i}{N}\bm{\sigma}_3(\lambda_\epsilon \bm{I}+i\lambda\bm{\sigma}_3\bm{L})^{-1}\right)\exp\left(-N\left\{\text{tr}\left(\frac{\lambda^2}{4J^2}\bm{L}^2-\frac{1}{2}\log\left(\lambda_\epsilon\bm{I}+i\lambda\bm{\sigma}_3\bm{L}\right)\right)\right\}\right)\,.
\end{equation}

\begin{figure}
    \centering
    \begin{minipage}{0.48\textwidth}
        \centering
        \includegraphics[width=\textwidth]{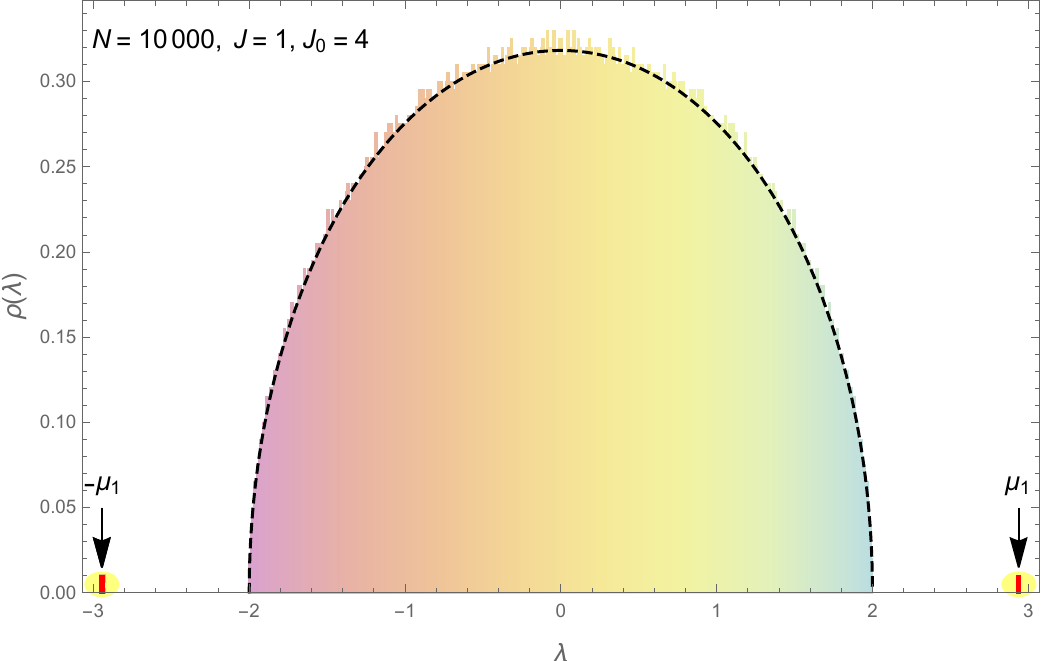}
    \end{minipage}\hfill
    \begin{minipage}{0.48\textwidth}
        \centering
        \includegraphics[width=\textwidth]{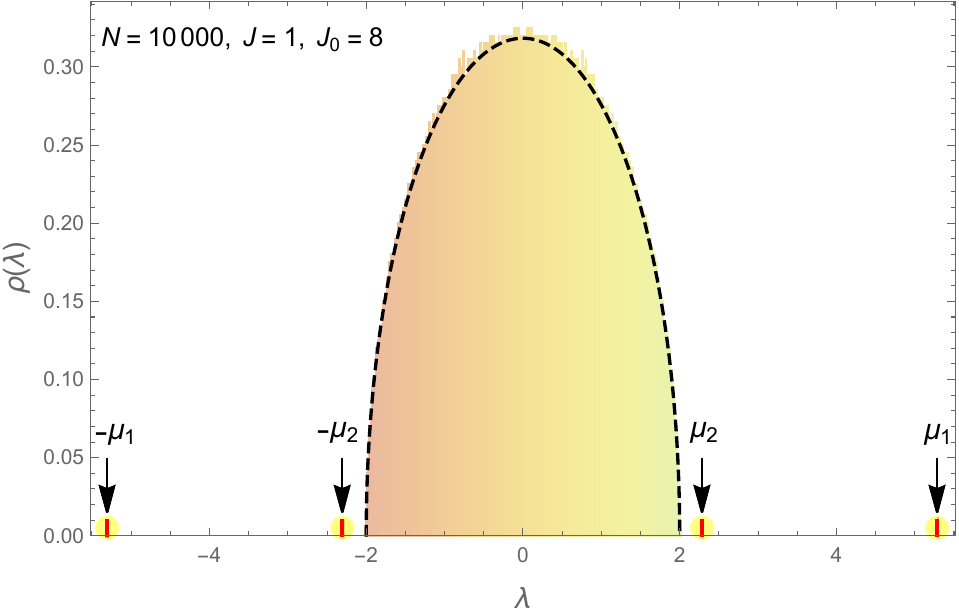}
    \end{minipage}
    \caption{Wigner semicircle and emerging delta peaks. The dotted line corresponds $\rho_0(\lambda).$}
    \label{Anti}
\end{figure}

Let us use again the method of steepest descent. It should be noted that prefactor
\begin{equation}
    \prod\limits_{i=1}^{N}\text{det}^{\frac{1}{2}}\left(\bm{I}+\frac{J_0\lambda_i}{N}\bm{\sigma}_3(\lambda_\epsilon \bm{I}+i\lambda\bm{\sigma}_3\bm{L})^{-1}\right)
\end{equation}
does not influence the saddle-point equations.
\begin{equation}
    \rho_{J_0}(\lambda)=\frac{1}{\pi N}\textrm{Im} \frac{\partial}{\partial \lambda}\left(\sum_{k=1}^{N}\log\left(1-\frac{J_0\lambda_k}{N(\lambda+i\lambda x^--i\epsilon)}\right)-Ng(x^-)\right)\,.
\end{equation}
Substituting $x^-$, we obtain
\begin{equation}
    \rho_{J_0}(\lambda)=\rho_0(\lambda)+\frac{1}{\pi N}\textrm{Im} \frac{\partial}{\partial \lambda}\sum_{k=1}^{N}\log\left(1- \frac{J_0\lambda_k}{2J^2N}(\lambda+i\sqrt{4J^2-\lambda^2})-iu_k\epsilon\right),
\end{equation}
where $u_k(\lambda)=\frac{J_0\lambda_k}{N\lambda^2(1+ix^-)^2}$. Without loss of generality, let $\lambda>0$. We know that $\rho_0(\lambda)=\rho_0(-\lambda)$ due to the matrix antisymmetry. We consider 2 cases:
\begin{itemize}
    \item $\lambda>2J$.
    \begin{equation}
    \rho_{J_0}(\lambda)=\rho_0(\lambda)+\frac{1}{\pi N}\textrm{Im} \frac{\partial}{\partial \lambda}\sum_{k=1}^{N}\log\left(1-\frac{J_0\lambda_k}{2J^2N}(\lambda-\sqrt{\lambda^2-4J^2})-iu_k\epsilon\right).
\end{equation}
We define a temporary function $f(\lambda)=1-\frac{J_0\lambda_k}{2J^2N}(\lambda-\sqrt{\lambda^2-4J^2})$. Expand $f(\lambda)$ in the Taylor series near its zero $f(\mu_k)=0$, where $\mu_k=\lambda^*_k+\frac{J^2}{\lambda^*_k}$ and $\lambda^*_k=\frac{J_0}{N}\lambda_k$:
\begin{equation}
    f(\lambda)=f'(\mu_k)(\lambda-\mu_k)+\mathcal{O}((\lambda-\mu_k)^2),\quad f'(\mu_k)=\frac{NJ_0\lambda_k}{J_0^2\lambda_k^2-J^2N^2}\,,
\end{equation}
\begin{equation}
    \frac{\partial}{\partial\lambda}\log(f'(\mu_k)(\lambda-\mu_k)+\mathcal{O}((\lambda-\mu_k)^2)-iu_k\epsilon)=\frac{1-\frac{iu'_k(\lambda)\epsilon}{f'(\mu_k)}+\mathcal{O}(\lambda-\mu_k)}{\lambda-\mu_k-\frac{iu_k\epsilon}{f'(\mu_k)}+\mathcal{O}((\lambda-\mu_k)^2)}\,.
\end{equation}
We are interested in singular terms because of the factor $\frac{1}{N}$, which suppresses any regular term. Using the Sokhotski–Plemelj formula, we get
\begin{equation}\label{rhoJ0:0}
    \rho_{J_0}(\lambda)=\rho_0(\lambda)+\frac{1}{N}\sum\limits_{k}\delta(\lambda-\mu_k),\quad k\in\left\{j:f(\mu_j)=0\right\}
\end{equation}
Value $\mu_k$ can be a zero of the function $f(\lambda)$ only in several cases, depending on $J_0$, $J$, $N$. Let us investigate it. Notice that
\begin{equation}
    \sqrt{\lambda^2-4J^2}-\lambda>
    \begin{cases}
        \begin{aligned}
            &2J\,, &\quad &\lambda<-2J\,;\\
            &-2J\,, &\quad &\lambda>2J\,.
        \end{aligned}
    \end{cases}
\end{equation}
Let us simplify the value of the $f(\lambda)$. This function can vanish in several cases:
\begin{equation}
    \begin{cases}
        \begin{aligned}
            &2J^2N<2JJ_0\lambda_k,&\quad &\lambda>2J,\;\lambda_k>0\,;\\
            &2J^2N>-2JJ_0\lambda_k,&\quad &\lambda<-2J,\;\lambda_k<0\,.
        \end{aligned}
    \end{cases}
\end{equation}
We will remain from (\ref{rhoJ0:0}) $\delta$-functions with $k$ which satisfy $\frac{J_0}{N}\lambda_k>J$ and $-\frac{J_0}{N}\lambda_k<J$, which is trivial.
\begin{equation}\label{rhoJ0:2}
    \rho_{J_0}(\lambda)=\rho_0(\lambda)+\frac{1}{N}\sum_{k}\delta(\lambda-\mu_k),\quad k\in\left\{j:\lambda^*_j>J\right\}\,.
\end{equation}
\item $\lambda<2J$.\\
In this case we only have a regular term.
\begin{equation}
    \Im\frac{\partial}{\partial \lambda}\sum_{k=1}^{N}\log\left(1- \frac{J_0\lambda_k(\lambda+i\sqrt{4J^2-\lambda^2})}{2J^2N}\right)=\sum\limits_{k=1}^N\frac{J_0\lambda_{k}}{\sqrt{4J^2-\lambda^2}}\frac{\lambda N/2-J_0\lambda_k}{J^2_0\lambda^2_k-J_0\lambda_k\lambda N+J^2N^2}\,.
\end{equation}
In the limit $N\gg1$ we get
\begin{equation}
    \rho_{J_0}(\lambda)=\rho_0(\lambda)\,.
\end{equation}
\end{itemize}
Considering the spectrum is even and combining the results of both cases, we get 
\begin{equation}\label{rhoJ0:3}
    \rho_{J_0}(\lambda)=\rho_0(\lambda)+\frac{1}{N}\sum\limits_{k\in\left\{j:\lambda^*_j>J\right\}}\delta(\lambda-\mu_k)+\delta\left(\lambda+\mu_k\right)\,.
\end{equation}
This result corresponds with numerical simulation from Wolfram Mathematica 13.0. We can see this in fig. \ref{Anti}.

Thus there are a set of critical values $J^{crit}_{0,k}=\frac{JN}{\lambda_{k}},\; k\in\{1,...,N/2\}.$ When the value of $J_0$ passes the critical value $J^{crit}_{0,k},$ a two new delta peaks $\delta(\lambda\pm\mu_k)$ appear. In order to make the form of the answer record similar to the answer in the case of a symmetric matrix, let us separately identify the smallest among the critical values $J^{crit}_0=J^{crit}_{0,1}=JN\tan{\frac{\pi}{2N}}\underset{N\rightarrow\infty}{=}\frac{\pi}{2}J$ below which no effect occurs. We found that the value of $J_0$ can significantly affect the properties of the spectrum. A phase transition occurs: depending on $J_0$, we can have a separation of the largest eigenvalues. We rewrite (\ref{rhoJ0:3}) in a more comprehensible way:
\begin{equation}\label{rhoJ0:4}
    \rho_{J_0}(\lambda)\equiv\braket{\nu(\lambda)}_{J_{ij}:J_0\neq0}=
    \begin{cases}
        \begin{aligned}
            &\rho_0(\lambda), &\quad &J_0<J^{crit}_{0,1}\,;\\
            &\rho_0(\lambda)+\frac{1}{N}(\delta(\lambda-\mu_{1})+\delta(\lambda+\mu_{1})), &\quad &J^{crit}_{0,1}<J_0<J^{crit}_{0,2}\,;\\
            & \hdots \\
            &\rho_0(\lambda)+\frac{1}{N}\sum_{k=1}^{N/2}\delta(\lambda-\mu_k)+\delta\left(\lambda+\mu_k\right), &\quad &J^{crit}_{0,\frac{N}{2}}<J_0\,.
        \end{aligned}
    \end{cases}
\end{equation}

Let us notice interesting fact. We present the antisymmetric matrix with a non-zero mean as the sum of the non-zero mean matrix (with the Wigner semicircle spectrum) and the constant matrix (with a spectrum of $\lambda_k$). And the spectrum of the sum of these two matrices is similar to the superposition of two spectra but there are several conditions on the appearance of the single states (\ref{rhoJ0:4}) and delta-peaks are shifted further away from the semicircle in comparison with constant matrix ($\mu_k$ instead of $\lambda_k$). Let us call this effect "spectrum interference". The same situation was in the case of symmetric matrices in \cite{Edwards_1976,Jones_1978}.

\section{Acknowledgements}
The authors are grateful to acknowledge helpful discussions with A. Litvinov and A. Kamenev. Special thanks to G. Tarnopolsky, who inspired us to solve this problem and helped during the writing of this article. This work has been supported by the Russian Science Foundation under the grant 22-22-00991.

\appendix
\section{Spectrum of the constant antisymmetric matrix}\label{ConstantMatrix}
Let us consider a constant antisymmetric matrix $A_{ij}$ of size $N$:
\begin{equation}
\bm{A} = \left(\begin{array}{ccccc}
    0 & 1 & \dots & 1 & 1 \\ 
    -1 & 0 & \dots & 1 & 1 \\ 
    \vdots & \vdots & \ddots & \vdots & \vdots \\
    -1 & -1 & \dots & 0 & 1 \\ 
    -1 & -1 & \dots & -1 & 0 \\ 
  \end{array}\right)\,.
\end{equation}
We obtain the expression of the characteristic polynomial using linear transformations for the determinant
\begin{equation}
    \det (\bm{A}-\lambda\bm{I}) = \underbrace{\left|
    \begin{array}{cccc}
	-\lambda & 1 & \ldots & 1\\
	-1 & -\lambda& \ldots & 1\\
	\vdots & \vdots & \ddots & \vdots\\
	-1 & -1 & \ldots & -\lambda
    \end{array}\right|}_{D_N(\lambda)}  =  -2\lambda D_{N-1}(\lambda)-(\lambda^2-1)D_{N-2}(\lambda) \,.
\end{equation}
The characteristic polynomial
\begin{equation}
    D_N(\lambda)=-2\lambda D_{N-1}(\lambda)-(\lambda^2-1)D_{N-2}(\lambda)\,,
\end{equation}
has boundary conditions $D_1=-\lambda,\;D_2=\lambda^2+1$. Solving the recurrent equation, we obtain
\begin{equation}
    D_N(\lambda)=\frac{1}{2}(-\lambda+1)^N+\frac{1}{2}(-\lambda-1)^N\,.
\end{equation}
Therefore, the eigenvalues have the following form
\begin{equation}
    i\lambda_k = i \cot{\left(\pi\frac{2k-1}{2N}\right)}, \; k\in \{1,2,..,N\}\,.
\end{equation}
The set $\lambda_k$ is ordered in descending order.
\section{Cartan classes}\label{Cartan}
Cartan classification is based on the fact that all random matrices can be divided into 10 symmetrical classes (see table \ref{tab:Cartan}). The set of 10 classes corresponds to a special set of manifolds known as Riemannian symmetric spaces, which have exceptional symmetry properties. For example, you can see \cite{2008}.

\begin{table}[ht]
    \centering
    \renewcommand{\arraystretch}{1.5}
    \begin{tabular}{|c|c|c|c|c|}
        \hline
        \textbf{Class} & \textbf{T} & \textbf{P} & \textbf{S} & \textbf{Structure} \\
        \hline
        A "Unitary" & $0$ & $0$ & $0$ & $\mathfrak{u}(N)$ \\
        \hline
        D & $0$ & $1$ & $0$ & $\mathfrak{so}(2N)$ \\
        \hline
        C & $0$ & $-1$ & $0$ & $\mathfrak{sp}(2N)$ \\
        \hline
        AI "Orthogonal" & $1$ & $0$ & $0$ & $\mathfrak{u}(N)/\mathfrak{so}(N)$ \\
        \hline
        BDI & $-1$ & $0$ & $0$ & $\mathfrak{so}(2N)/\mathfrak{so}(N)\times \mathfrak{so}(N)$ \\
        \hline
        CI & $1$ & $-1$ & $1$ & $\mathfrak{sp}(2N)/\mathfrak{u}(N)$ \\
        \hline
        AII "Symplectic" & $-1$ & $0$ & $0$ & $\mathfrak{u}(2N)/\mathfrak{sp}(2N)$ \\   \hline
        DIII & $-1$ & $1$ & $1$ & $\mathfrak{so}(2N)/\mathfrak{u}(N)$ \\
        \hline
        CII & $-1$ & $-1$ & $1$ & $\mathfrak{sp}(4N)/\mathfrak{sp}(2N)\times \mathfrak{sp}(2N)$ \\
        \hline
        AIII & $0$ & $0$ & $1$ & $\mathfrak{u}(2N)/\mathfrak{u}(N)\times \mathfrak{u}(N)$ \\
        \hline
    \end{tabular}
    \caption{The Cartan Classification of Topological Insulators.}
    \label{tab:Cartan}
\end{table}

Consider the model of spinless non-interacting fermions on a lattice with a Hamiltonian:
\begin{equation}\label{spinless}
    H=\sum\limits_{i,j}t_{ij}c_i^\dagger c_i+\frac{1}{2}\Delta_{ij}c_i^\dagger c_j^\dagger+\frac{1}{2}\Delta^\dagger_{ij}c_ic_j\,,
\end{equation}
where matrix $t_{ij}$ is Hermitian $\bm{t}^\dagger=\bm{t}$.

It encodes how electrons “tunnel” from site to site on the lattice. The anomalous terms are proportional to $\Delta$ and $\Delta^\dagger$, where $\Delta_{ij}$ is the superconducting gap parameter. A normal metal or insulator has $\Delta = \Delta^\dagger = 0$.\\
It is convenient to rewrite Hamiltonian in a different way. Introduce the Majorana spinor:
\begin{equation}
    \chi\equiv\begin{pmatrix}
        c\\
        (c^\dagger)^T
    \end{pmatrix}\Leftrightarrow\chi_{i,\sigma}=\begin{cases}
        c_i\,,\quad\sigma=1\,;\\
        c^\dagger_i\,,\quad\sigma=2\,,
    \end{cases}
\end{equation}
where $i\in\{1,...,N\}$ is a site index and $\sigma\in\{1,2\}$ is a "particle-hole" index.
The Hermitian conjugate quantity $\chi^\dagger$ is independent of $\chi$
\begin{equation}
    \chi^\dagger=\begin{pmatrix}
        c^\dagger & c^T
    \end{pmatrix}=\chi^T\bm{\sigma}_1\,.
\end{equation}
Pauli $2N\times2N$ matrices:
\begin{equation}
    \bm{\sigma}_1=\begin{pmatrix}
        0 & \bm{I}\\
        \bm{I} & 0
    \end{pmatrix},\quad\bm{\sigma}_2=\begin{pmatrix}
        0 & -i\bm{I}\\
        i\bm{I} & 0
    \end{pmatrix},\quad\bm{\sigma}_3=\begin{pmatrix}
        \bm{I} & 0\\
        0 & -\bm{I}
    \end{pmatrix}\,.
\end{equation}
We can rewrite the Hamiltonian (\ref{spinless}) as
\begin{equation}
    H=\frac{1}{2}\chi^T\bm{M}_P\bm{h}\chi,\quad\bm{M}_P=\bm{\sigma}_1,\quad\bm{h}=\begin{pmatrix}
        \bm{t} & \bm{\Delta}\\
        \bm{\Delta}^\dagger & -\bm{t}^T
    \end{pmatrix},
\end{equation}
where the Hermitian matrix $\bm{h}=\bm{h}^\dagger$ is the effective single-particle Hamiltonian. In general case $\bm{M}_P$ is not necessarily $\bm{\sigma}_1$.

Now let us define symmetry conditions on $\bm{h}$. We do not consider spatial symmetries (translations and rotations) because we apply this theory to random matrices.  
\begin{itemize}
    \item Particle-hole symmetry $P$
    \begin{equation}
        \bm{M}_P\bm{h}^T\bm{M}_P=-\bm{h},\quad\bm{M}_P=\begin{cases}
            +\bm{M}^T_P\,,\quad P^2=1\,;\\
            -\bm{M}^T_P\,,\quad P^2=-1.
        \end{cases}
    \end{equation}
    For example, $\bm{M}_P=\bm{\sigma}_1$ -- symmetric matrix and it corresponds to $P^2=1$.
    \item Time-reversal invariance $T$
    \begin{equation}
        \bm{M}_T\bm{h}^*\bm{M}_T=\bm{h},\quad\bm{M}_T=\begin{cases}
            +\bm{M}^T_T\,,\quad T^2=1\,;\\
            -\bm{M}^T_T\,,\quad T^2=-1\,.
        \end{cases}
    \end{equation}
    \item Chiral (sublattice) symmetry $S$ 
    \begin{equation}
        -\bm{M}_S\bm{h}\bm{M}_S=\bm{h},\quad\bm{M}_S=\bm{M}^\dagger_S=\bm{M}^{-1}_S\,.
    \end{equation}
\end{itemize}
\bibliography{Refs.bib}

\begin{thebibliography}{8}
\expandafter\ifx\csname natexlab\endcsname\relax\def\natexlab#1{#1}\fi
\expandafter\ifx\csname bibnamefont\endcsname\relax
  \def\bibnamefont#1{#1}\fi
\expandafter\ifx\csname bibfnamefont\endcsname\relax
  \def\bibfnamefont#1{#1}\fi
\expandafter\ifx\csname citenamefont\endcsname\relax
  \def\citenamefont#1{#1}\fi
\expandafter\ifx\csname url\endcsname\relax
  \def\url#1{\texttt{#1}}\fi
\expandafter\ifx\csname urlprefix\endcsname\relax\def\urlprefix{URL }\fi
\providecommand{\bibinfo}[2]{#2}
\providecommand{\eprint}[2][]{\url{#2}}

\bibitem[{\citenamefont{Mehta}(2004)}]{Meh2004}
\bibinfo{author}{\bibfnamefont{M.~L.} \bibnamefont{Mehta}},
  \emph{\bibinfo{title}{Random Matrices}} (\bibinfo{year}{2004}),
  \bibinfo{edition}{3rd} ed.

\bibitem[{\citenamefont{Mehta and Rosenzweig}(1968)}]{MEHTA1968449}
\bibinfo{author}{\bibfnamefont{M.}~\bibnamefont{Mehta}} \bibnamefont{and}
  \bibinfo{author}{\bibfnamefont{N.}~\bibnamefont{Rosenzweig}},
  \bibinfo{journal}{Nuclear Physics A} \textbf{\bibinfo{volume}{109}},
  \bibinfo{pages}{449} (\bibinfo{year}{1968}), ISSN \bibinfo{issn}{0375-9474},
  \urlprefix\url{https://www.sciencedirect.com/science/article/pii/0375947468906118}.

\bibitem[{\citenamefont{Edwards and Jones}(1976)}]{Edwards_1976}
\bibinfo{author}{\bibfnamefont{S.~F.} \bibnamefont{Edwards}} \bibnamefont{and}
  \bibinfo{author}{\bibfnamefont{R.~C.} \bibnamefont{Jones}},
  \bibinfo{journal}{Journal of Physics A: Mathematical and General}
  \textbf{\bibinfo{volume}{9}}, \bibinfo{pages}{1595} (\bibinfo{year}{1976}),
  \urlprefix\url{https://doi.org/10.1088/0305-4470/9/10/011}.

\bibitem[{\citenamefont{Jones et~al.}(1978)\citenamefont{Jones, Kosterlitz, and
  Thouless}}]{Jones_1978}
\bibinfo{author}{\bibfnamefont{R.~C.} \bibnamefont{Jones}},
  \bibinfo{author}{\bibfnamefont{J.~M.} \bibnamefont{Kosterlitz}},
  \bibnamefont{and} \bibinfo{author}{\bibfnamefont{D.~J.}
  \bibnamefont{Thouless}}, \bibinfo{journal}{Journal of Physics A: Mathematical
  and General} \textbf{\bibinfo{volume}{11}}, \bibinfo{pages}{L45}
  (\bibinfo{year}{1978}),
  \urlprefix\url{https://doi.org/10.1088/0305-4470/11/3/002}.

\bibitem[{\citenamefont{Klebanov and Tarnopolsky}()}]{KlebTarn}
\bibinfo{author}{\bibfnamefont{I.}~\bibnamefont{Klebanov}} \bibnamefont{and}
  \bibinfo{author}{\bibfnamefont{G.}~\bibnamefont{Tarnopolsky}},
  \bibinfo{note}{unpublished}.

\bibitem[{\citenamefont{Ivanov}(2002)}]{2002}
\bibinfo{author}{\bibfnamefont{D.~A.} \bibnamefont{Ivanov}},
  \bibinfo{journal}{Journal of Mathematical Physics}
  \textbf{\bibinfo{volume}{43}}, \bibinfo{pages}{126–153}
  (\bibinfo{year}{2002}), ISSN \bibinfo{issn}{1089-7658},
  \urlprefix\url{http://dx.doi.org/10.1063/1.1423765}.

\bibitem[{\citenamefont{Kamenev and M\'{e}zard}(1999)}]{Kamenev_1999}
\bibinfo{author}{\bibfnamefont{A.}~\bibnamefont{Kamenev}} \bibnamefont{and}
  \bibinfo{author}{\bibfnamefont{M.}~\bibnamefont{M\'{e}zard}},
  \bibinfo{journal}{Journal of Physics A: Mathematical and General}
  \textbf{\bibinfo{volume}{32}}, \bibinfo{pages}{4373} (\bibinfo{year}{1999}),
  \urlprefix\url{https://doi.org/10.1088%2F0305-4470%2F32%2F24%2F304}.

\bibitem[{\citenamefont{Schnyder et~al.}(2008)\citenamefont{Schnyder, Ryu,
  Furusaki, and Ludwig}}]{2008}
\bibinfo{author}{\bibfnamefont{A.~P.} \bibnamefont{Schnyder}},
  \bibinfo{author}{\bibfnamefont{S.}~\bibnamefont{Ryu}},
  \bibinfo{author}{\bibfnamefont{A.}~\bibnamefont{Furusaki}}, \bibnamefont{and}
  \bibinfo{author}{\bibfnamefont{A.~W.~W.} \bibnamefont{Ludwig}},
  \bibinfo{journal}{Physical Review B} \textbf{\bibinfo{volume}{78}}
  (\bibinfo{year}{2008}), ISSN \bibinfo{issn}{1550-235X},
  \urlprefix\url{http://dx.doi.org/10.1103/PhysRevB.78.195125}.

\end{thebibliography}
\end{document}